\documentclass{aastex}          
\usepackage{spr-astr-addons}    

\begin{document}
%
  \title{Properties of the narrow line Seyfert 1 galaxies revisited}

\shorttitle{Properties of the narrow line Seyfert 1 galaxies revisited}

\shortauthors{Liu et al.}

\author{Xiang Liu\altaffilmark{1,2}}
\and
\author{Pingping Yang\altaffilmark{1,3}}
\and
\author{Renzhi Su\altaffilmark{1,3}}
\and
\author{Zhen Zhang\altaffilmark{1}}

\email{liux@xao.ac.cn}

\altaffiltext{1}{Xinjiang Astronomical Observatory, Chinese
Academy of Sciences, 150 Science 1-Street, Urumqi 830011, PR
China}

\altaffiltext{2}{Key Laboratory of Radio Astronomy, Chinese
Academy of Sciences, Urumqi 830011, PR China}

\altaffiltext{3}{Graduate University of Chinese Academy of
Sciences, Beijing 100049, PR China}

\begin{abstract}

There is growing evidence to suggest that the black hole mass has been previously underestimated with the H$\beta$ line width for certain active galactic nuclei (AGN). With the assumption of the flatter rather than isotropic velocity distribution of gases in the broad-line region of AGN, we investigated the properties of narrow line Seyfert 1 (NLS1) galaxies, like the black hole mass and the Eddington ratio, and compared with broad line Seyfert 1 (BLS1) galaxies. Since gamma-rays detected in a few NLS1s which favor a smaller viewing angle in NLS1s than BLS1s, with the projection effect we estimated the relative black hole mass and Eddington ratio for NLS1s and BLS1s. The result implies that the NLS1s and BLS1s have similar black hole masses and Eddington ratios, peaked at a larger black hole mass and lower Eddington ratio for the NLS1s than thought before. Furthermore, with applying the correction factor 6 of average black hole mass as derived from the modelling of both optical and UV data in radio-loud NLS1s by Calderone et al., to the Xu et al. sample, we find that the NLS1s and BLS1s also show similar black hole masses and Eddington ratios, peaked at $2.0\times10^{7}M_{\odot}$ and 0.12 (Eddington ratio) for the NLS1s. The $M_{BH}-\sigma$ relation due to the enhanced black hole masses of NLS1s is discussed. In addition, there seems to show a linear correlation between jet power and disk luminosity for the flat spectrum radio-loud NLS1 sample, which implies an accretion dominated rather than black hole spin dominated jet.

\end{abstract}


\section{Introduction}

Since the detection of GeV gamma-ray emission in some of narrow line Seyfert 1 (NLS1) galaxies with the Fermi-LAT, this kind of AGN becomes a hot topic in the AGN research field. The `narrow line Seyfert 1' was first denoted by Osterbrock \& Pogge (1985), who found that the NLS1s have properties that different to normal Seyfert 1 galaxies, as quantified by Goodrich (1989): 1) unusually narrower HI lines, i.e. the Balmer lines are only slightly broader than forbidden lines such as [O III], [N II] and [S II], with a quantified criterion of FWHM$(H\beta) < 2000\,km/s$; 2) the line ratio [O III] $\lambda 5007/H\beta<3$, a level found discriminated well between Seyfert 1s and 2s; 3) there are often present emission lines from FeII or higher ionization iron lines. The first criterion, as noted by Goodrich (1989), that of line width is similar to that used by Khachikian \& Weedman (1974) in their original definition of the Seyfert 1 and 2 classes and hence many of the narrow line 1s have been classified at times as Seyfert 2s. The last criterion indicates a closer relationship to Seyfert 1s, since Seyfert 2s do not show strong FeII and in general do not show strong high-ionization iron lines (Goodrich 1989).

Because gamma-rays are detected in a few NLS1s with the Fermi-LAT, and some of the radio loud NLS1s have shown blazar's properties (Yuan et al. 2008), it is possible that the radio loud NLS1s are mostly pole-on type 1 AGN. However, the question which arises about the unusually narrow Balmer lines is still not well understood. It is proposed that the broad line region (BLR) is likely more `flat' than a spherical/isotropic structure (Decarli et al. 2008; Stern \& Poutanen 2014), which would shed light on the issue of `unusually narrow H$\beta$ line' in the NLS1s. Recently, it is suggested that the black hole mass of some NLS1s could be much larger than estimated before (Calderone et al. 2013; Baldi et al. 2016), this will have significant implications on the properties of NLS1s we thought before such as: small black hole ($10^{6.5}$ solar mass) and very high Eddington ratio (close to 1), see e.g., Komossa et al. (2006), Zhou et al. 2006, Xu et al. (2012). We intend to revisit the properties of NLS1s with the flat BLR model and new findings of black hole mass, in this paper.

\section{The `flat' BLR of AGN and its implications}

It is reasonable to assume that the AGN in NLS1s are more face-on to us than in the broad line Seyfert 1 galaxies (BLS1s), because the former has a higher gamma-ray detection rate than the latter (Ackermann et al. 2015). The radio loud and/or the gamma-ray loud NLS1s are likely pole-on to us like blazars with the jet viewing angle of less than few degrees, and the radio loud NLS1s (RL-NLS1s) could be the low-mass/high-accretion tail of blazars distribution (Foschini et al. 2015). The pole-on RL-NLS1s will naturally lead to the `narrower' broad emission lines, assuming that the broad line region is flatter than spherical/isotropic geometry as suggested by Decarli et al. (2008). Recently, with the gamma-ray absorption studies through the $\gamma-\gamma$ pair creation and breaks in the GeV gamma-ray spectra, the high energy gamma-rays of blazars are thought to come from a flat disk-like BLR (Tavecchio \& Ghisellini 2012; Lei \& Wang 2014; Stern \& Poutanen 2014).
The flatter BLR structure will have less opacity for gamma-rays, and causes the narrower width of broad emission lines of NLS1s, this is attributed to a viewing angle effect in which most of the motion of the BLR clouds is constrained to a disk like structure which perpendicular to our line of sight (Orr \& Browne 1982; Boroson \& Green 1992).

Due to the selection effect by definition, the
NLS1s have narrow broad line width of H$\beta < 2000\,km/s$ than the BLS1s (H$\beta \geq 2000\,km/s$), so that the NLS1s will have smaller viewing angles than the BLS1s in the light of the flat BLR scenario. In this explanation of the unusually narrow broad line of NLS1s, the real virial velocity of the BLR clouds should be much larger than the observed narrow broad lines, i.e. the observed velocity of narrow broad line is actually a fraction of the real velocity of the BLR clouds due to projection effect. This will have important implications, for instance, on the black hole mass and the Eddington ratio estimated with H$\beta$ line. The black hole mass would be underestimated with the narrow broad line velocity and the Eddington ratio then is overestimated (Eddington luminosity is proportional to the black hole mass) for the NLS1s.

\section{BH mass and Eddington ratio of NLS1s}

The black hole mass at the center of NLS1s is usually estimated using the velocity of H$\beta$ which is assumed to be the virial velocity of the BLR clouds, e.g. in Bentz et al. (2009), Xu et al. (2012), but as noted in the previous section the black hole mass measured with the H$\beta$ for NLS1s is most likely underestimated, because the observed FWHM of the H$\beta$ is dependent on the orientation of the line of sight
to the gas of BLR. Black hole mass estimation with other emission lines or methods has to be tried and compared with the result of using H$\beta$.

Bian \& Zhao (2004) find the black hole mass of NLS1s estimated with [O III] is about 10 times larger than that estimated with H$\beta$, the [O III] could be less affected by the viewing angle than the H$\beta$, although they thought the result from the [O III] was less reliable from their samples. Bian \& Zhao (2004) also used the soft X-ray excess as a prominent character of NLS1s to estimate the black hole masses, the resulted black hole mass of NLS1s are also larger on average than that from H$\beta$. Nikolajuk et al. (2009) generalized the mass determination method
based on the X-ray excess variance, find that the black hole mass is about 2 times larger than that from H$\beta$. Calderone et al. (2013) used a new method to estimate black hole masses which relies on the modelling of both optical and UV data with a Shakura
\& Sunyaev disc spectrum, to a sample of 23 radio-loud NLS1 galaxies, find that the black hole masses are consistently larger than $10^{8}M_{\odot}$ at least a factor of 6 above previous results based on single epoch virial methods, while the Eddington ratios are correspondingly lower. Decarli et al. (2008) considered the geometrical factor relating the observed width (FWHM) of broad lines to the virial mass of black hole, assuming a flat BLR in NLS1s, they
show that the geometrical factor can fully account for the ¡®black hole mass deficit¡¯ observed in
NLS1s, and that Eddington ratio is (on average) comparable to the value of the more common broad-line
Seyfert 1 galaxies. Recently, Baldi et al. (2016) find that for RL-NLS1 PKS 2004$-$447, a polarized H$\alpha$ line with a width of 9000 km/s, 6 times broader than the width seen in direct light, this corresponds to a revised estimate
of black hole mass $6\times10^{8}M_{\odot}$, much higher than the $5\times10^{6}M_{\odot}$ estimated with H$\beta$.

The growing evidence suggests that the H$\beta$ line could be not good to be used for the estimate of black hole mass in NLS1s, but it may be a good indicator of the viewing angle of radio jet axis (Punsley \& Zhang 2010). If we re-scale for the black hole masses in Xu et al. (2012) with the correction factor of 6 as derived from 23 radio loud NLS1s in Calderone et al. (2013), we can see that the NLS1s and BLS1s have similar range of black hole masses as well as the Eddington ratios in Fig.1 and Fig.2, which peaks at $\sim 2.0\times10^{7}M_{\odot}$ and 0.12 (Eddington ratio) respectively for the NLS1s, and peaks at $\sim 1.4\times10^{7}M_{\odot}$ and 0.18 (Eddington ratio) for the BLS1s. There is no significant difference found for the BH mass distributions, but different distributions found for the Eddington ratios from the K-S test. The remaining differences of the result between the NLS1s and BLS1s, could be due to that the mass correction has been made only for the NLS1s but not for the BLS1s.

\begin{figure}
    \includegraphics[width=8cm]{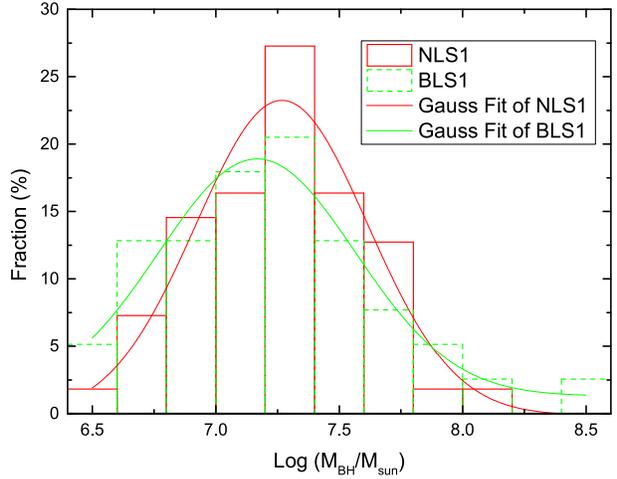}
    \caption{The distributions of the black hole masses of the NLS1s and BLS1s from Xu et al. (2012), in which the black hole mass of NLS1s has been multiplied by a factor of 6 taken from Calderone et al. (2013), and the corresponding Gaussian fittings.}
     \label{fig1}
  \end{figure}

\begin{figure}
    \includegraphics[width=8cm]{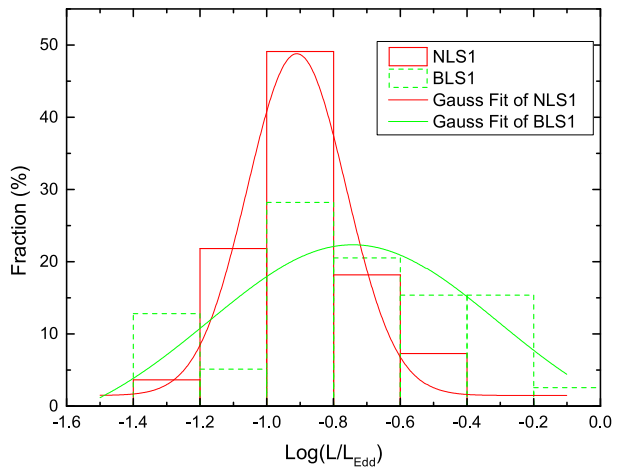}
    \caption{The distributions of the Eddington ratios of the NLS1s and BLS1s from Xu et al. (2012), in which the Eddington ratio of NLS1s has been divided by a factor of 6 taken from Calderone et al. (2013), and the corresponding Gaussian fittings.}
     \label{fig2}
  \end{figure}

We could further study this issue with the data from the sample of Xu et al. (2012) and assuming a `flat' BLR structure that a typical BLR cloud is moving in a longitudinal direction of the flat BLR with a virialized velocity of $V=6220$ km/s taken from the maximum value in the sample of Xu et al. (2012). With the median velocity (H$\beta$) of 1460 km/s and 2890 km/s in Xu et al. (2012), the derived median viewing angle ($\theta$) is $13.6^{\circ}$ and $27.7^{\circ}$ for the NLS1s and BLS1s respectively, where the observed H$\beta$ velocity $V_{obs}=Vsin(\theta)$. We try to re-sclae the black hole mass in Xu et al. (2012) with the virial-based scaling relationship as follows, and the resulted distributions are shown in Fig.3 and Fig.4.

\begin{equation}
M_{bh}= fR_{BLR}V^{2}/G,
\end{equation}

where $R_{BLR}$ is the broad-line region scale radius, and $V$ is
the typical velocity of BLR clouds, $G$ is the gravitational constant, $f$ is a general geometrical function
which correct for the unknown structure and inclination to the line of sight (e.g., see Wang et al. 2014), while it is often set as a constant (e.g., Decarli et al. 2008, Bentz et al. 2009), but here the projection effect is considered as $V_{obs}=Vsin(\theta)$.

The relative black hole mass and Eddington ratio distribution illustrated in Fig.3 and Fig.4 have similar peaks for the NLS1s and BLS1s (with no significant difference found for the BH mass distributions, although different distributions found for the Eddington ratios, from the K-S test), with the peak values of $5.6\times10^{7}M_{\odot}$ and 0.04 (Eddington ratio) for the NLS1s. However, this estimation is depending on the assumed real velocity of the BLR clouds that the maximum observed value is used here.

\begin{figure}
    \includegraphics[width=8cm]{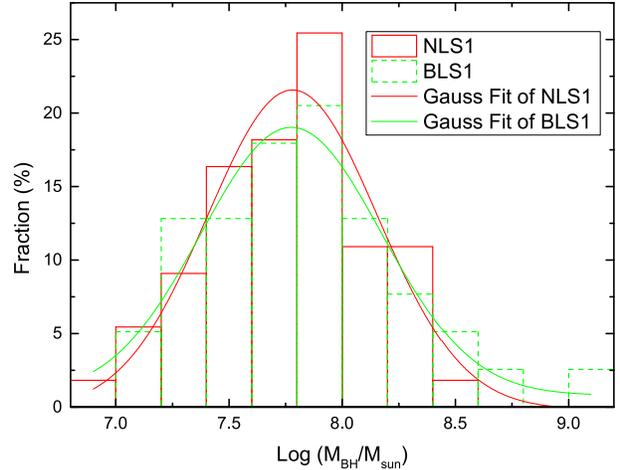}
    \caption{The distributions of the black hole masses of the NLS1s and BLS1s from Xu et al. (2012) with considering the effect of inclination angle, with the corresponding Gaussian fittings.}
     \label{fig3}
  \end{figure}

\begin{figure}
    \includegraphics[width=8cm]{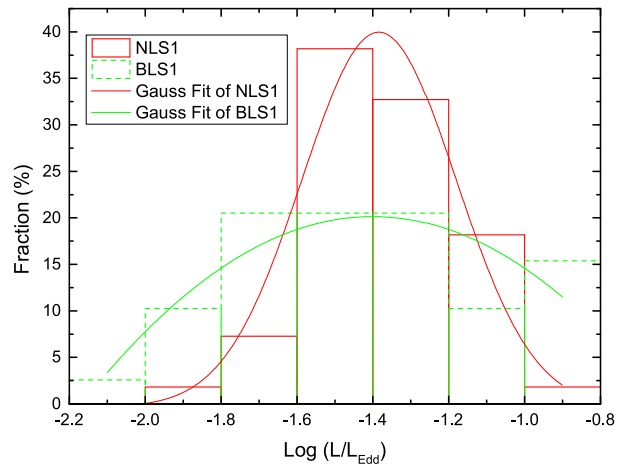}
    \caption{The distributions of the Eddington ratios of the NLS1s and BLS1s from Xu et al. (2012) with considering the effect of inclination angle, with the corresponding Gaussian fittings.}
     \label{fig4}
  \end{figure}

\section{Density of narrow-line region and $\gamma-rays$}

The BLR clouds have higher density than that of the narrow line region (NLR) clouds, with the electron density increasing inward as $n_{e}\propto r^{-2}$, $r$ is the distance from the central black hole, and the dispersion velocity of clouds in the BLR increases either with increasing $n_{e}$, or with increasing the ionization parameter which increases inward also $\propto r^{-2}$ (Osterbrock \& Mathews 1986). Xu et al. (2007, 2012) have discussed that the widths of [S II] doublet trace the stellar velocity
dispersion of galaxy bulge better than that of [O III], and they used the intensity ratio of the [S II] doublet ($\lambda6716/\lambda6731$) to
derive the density of NLR (in the range of $n_{e}\sim 10-10^{4}cm^{-3}$). Their result suggests that the BLS1s avoid low density, while the NLS1s show a wider distribution in the
NLR density, including a significant number of objects with low densities. There is tentative anti-correlation between the ratio of the [S II] doublet and the [S II] line velocity in the NLS1s and BLS1s from Xu et al. (2012), as the linear fitting shown in Fig.5. It is not clear the place of the emission region of [S II] clouds. If it is in the NLR, the projection effect could be not significant; if it is in the BLR, the [S II] line velocity of NLS1s could be affected by the projection effect, i.e. the [S II] line velocity would be underestimated for the NLS1s. In the latter case, the resultant NLR density of the NLS1s could be different than thought before, considering the anti-correlation between the ratio of the [S II] doublet and the [S II] line velocity of the NLS1s and BLS1s in Fig.5.

From the GeV gamma-ray breaks due to photon-photon pair production, the GeV gamma-rays are considered to come mainly from the BLR of blazars (Stern \& Poutanen 2014). However, not all known blazars have been detected in gamma-rays, some well known blazars still have no gamma-ray detection, e.g. the quasar 3C345, probably in these blazars their BLR density or opacity is so high that gamma-rays are largely absorbed through the photon-photon pair production and so become too faint to be detected by the Fermi-LAT threshold. The third catalog of active galactic nuclei detected by the Fermi-LAT contains 1563 AGN at high galactic latitude ($|b|>10^{\circ}$), of them 98\% are blazars, and 35 are non-blazar or misaligned AGN, including a few radio loud NLS1 galaxies (Ackermann et al. 2015; Paliya et al. 2015). Apart from the small viewing angles and relativistic jets the most important factors for the gamma-ray AGN, the BLR density may take a role for the opacity effect, the misaligned gamma-ray AGN may have smaller opacity in their BLR and NLR, so that the gamma-rays could go farther out the BLR to be detectable at larger viewing angles (e.g. Liao et al. 2015). In Xu et al. (2012), there are some outliers of NLS1s with much lower NLR density than the BLS1s, they would have a lower opacity for gamma-rays.

\begin{figure}
    \includegraphics[width=8cm]{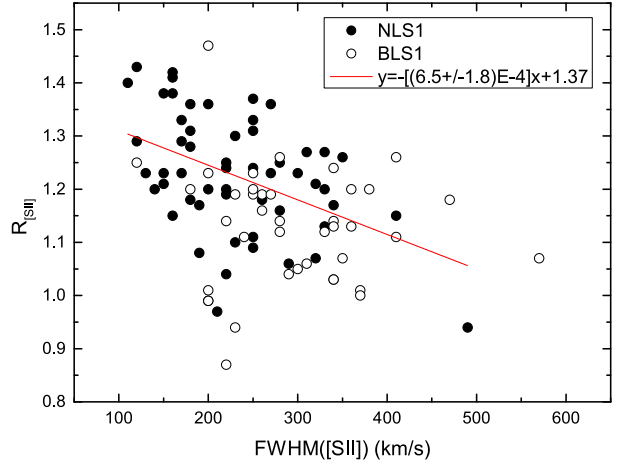}
    \caption{The ratio of [S II] doublet versus the FWHM ([S II]) of the NLS1 and BLS1 galaxies for the data of Xu et al. (2012), with the best linear fit.}
     \label{fig5}
  \end{figure}

  \begin{figure}
    \includegraphics[width=8cm]{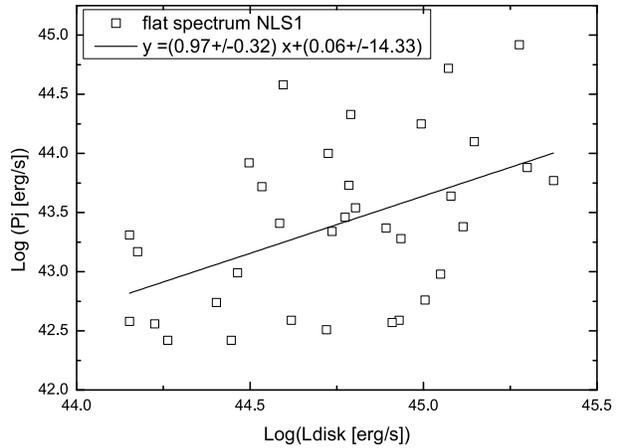}
    \caption{Radiative radio core power at 15 GHz versus disk luminosity for the flat spectrum radio-loud NLS1 galaxies of Foschini et al. (2015), with the best linear fit.}
     \label{fig6}
  \end{figure}

\section{Radio, X-ray properties and black hole spin}

Radio properties of NLS1s as summarised by Komossa et al. (2006), a fraction of radio loud NLS1s is 7\%, about a half of that for radio loud quasars. A few NLS1s have shown very high radio loudness ($>100$), implying strong relativistic jets (Zhou et al. 2003, Yuan et al. 2008), and later some of them are detected by Fermi-LAT in gamma-ray band. However, not all NLS1s have strong relativistic jets. Angelakis et al. (2015) find mild relativistic jets in NLS1s through multiwavelength flux variability. Doi et al. (2012) find that six NLS1s show kpc scale radio structure, including two gamma-ray sources with core-dominated, and the detection rate of extended radio emissions in NLS1s is lower than that in the broad-line AGN. And Gu et al. (2015) find from 14 NLS1s that 50\% have flat radio spectra, and another half shows steep spectra, implying some of NLS1s may have relatively larger jet viewing angles. Foschini et al. (2015) present the multiwavelength results of 42 flat spectrum radio-loud NLS1s, with 17\% having gamma-ray detection, and 90\% of the NLS1s are detected in X-rays, in which the X-ray spectral indices of the RL-NLS1s are
similar to those of BLS1s comparable to the result of Zhou \& Zhang (2010), and with usually harder spectra than those of radio quiet NLS1s. From X-ray Fe K$\alpha$ line, there are tentative indications that black holes in NLS1s may
not spin very fast with the dimensionless spin parameter $a<0.85$ on average (Liu et al. 2015).

Liu \& Han (2014) and Liu et al. (2016) studied the correlation between radio jet power and disk luminosity of AGN with the power-law correlation index of $\mu$, and there is a linear correlation for FRII quasars ($\mu\sim 1$) (van Velzen \& Falcke 2013) while the low luminosity AGN have flatter correlation indices ($\mu < 0.7$). They interpreted that the jet power is accretion dominated for the linear correlation and the black hole spin may play an important role in the flatter spectrum correlation. For the flat spectrum radio-loud NLS1s in Foschini et al. (2015), there seems to have a linear correlation between the jet power and disk luminosity after excluding few outliers in the low and high accretion luminosity. The result as shown in Fig.6, illustrates a slope of 0.97$\pm0.32$ for the radiative jet power (and a slope of 1.09$\pm0.36$ for the total jet power including kinetic power, not shown here), this is similar to the linear correlation of the gamma-ray loud blazars by Ghisellini et al. (2014), but with a larger scatter of fitting residuals. From the view of Liu et al. (2016) and also see Done et al. (2013), this result implies that the jet power in the RL-NLS1s could be dominated by the accretion disk rather than by the black hole spin.

\section{Discussion and summary}

It is reasonable to assume a flatter BLR to explain the different H$\beta$ line width, as also pointed out by Shen \& Ho (2015) for quasars in which orientation plays a significant role in determining the observed kinematics of the gas,
implying a flattened, disklike geometry for the fast-moving clouds
close to the black hole. The orientation angle could be correlated with the spectral index or radio core dominance for radio loud NLS1s, Brotherton et al. (2015) find that the FWHM of H$\beta$
correlates significantly with radio core dominance and biases black hole mass determinations
that use it, but that this is not the case for velocity dispersion $\sigma$ based on [O III] $\lambda$5007, by comparing the black hole mass estimated using H$\beta$ and [O III] $\lambda$5007 which tracks mass via the $M_{BH}-\sigma$ relation. However, whether the NLS1s follow the same $M_{BH}-\sigma$ relation built with bright AGN/quasars is still controversial, e.g. see Bian et al. (2008), Kormendy \& Ho (2013), Woo et al. (2015), and Shankar et al. (2016). Mathur et al. (2012) have showed results that the NLS1 galaxies are below the $M_{BH}-L_{bulge}$ relation, and this scaling relation can be caused by a combination of factors such as the galaxy morphology, orientation and evolution. If we take account of a flat BLR and the different orientation between NLS1s and BLS1s, as we analysed above, the mass of NLS1s can be heavier than that estimated with H$\beta$, so that the NLS1s could become closer to the $M_{BH}-L_{bulge}$ relation for their enhanced black hole masses.

As corrected with the orientation effect, the NLS1s and BLS1s may have similar black hole masses and comparable Eddington ratios. The detection of gamma-rays from some of radio loud NLS1s favors their smaller jet viewing angles than that of BLS1s. But other intrinsic factors may also play a role, e.g. the density of BLR and NLR of NLS1s, which could determine the gamma-ray opacity. Assuming a continuity of scaling as $n_{e} \propto r^{-2}$, some of the NLS1s in Xu et al. (2012) have much lower NLR density than that in BLS1s, and they could have less gamma-ray opacity. In addition, it is found that some `blue outliers' occur in the NLS1 galaxies with a fraction of 5-16\% (depending on the [O III] blue wing's velocity from 250 to 150 km/s by definition), but not found in the BLS1s (Komossa et al. 2008), suggesting that the outflows could be seen more often in the pole-on NLS1s than in the BLS1s, but the physical outflow in the NLR should be driven mainly by the accretion rate (Wang et al. 2016).

The possible linear correlation between the jet power and accretion luminosity of RL-NLS1s implies that the jet is the mass accretion dominated, as interpreted in Liu et al. (2016), and note that the Doppler beaming effect has been considered with the SED fitting program in Foschini (2014). The black hole seems not to spin very fast in the RL-NLS1s as estimated from the X-ray data (Liu et al. 2015), favoring the accretion dominated jet in the RL-NLS1s. The disk luminosities of the RL-NLS1s are largely above $10^{44}$ erg/s, greater than that of Seyfert galaxies and low luminosity AGN whose disk luminosity is mostly less than $10^{44}$ erg/s in Liu \& Han (2014), and the correlation index between the jet power and disk luminosity is steeper in the RL-NLS1s than that in the Seyferts and low luminosity AGN. However, the present sample of the RL-NLS1s is still quite small, we may need a larger sample of RL-NLS1s to confirm the result of linear correlation, and further multi-wavelength observation and model fitting are also required to study whether the black hole spins of radio loud NLS1s are indeed low/intermediat or not.

\section*{Acknowledgments}

This work is supported by the 973 Program 2015CB857100; the Key Laboratory of Radio Astronomy, Chinese Academy of Sciences; and the National Natural Science Foundation of China (No.11273050).

\end{document}